\documentstyle[preprint,tighten,aps,epsf,floats]{revtex}

\begin{document} 
 
\preprint{ 
\vbox{\hbox{CLNS-00/1707}\hbox{JHU--TIPAC--20007} 
\hbox{hep-ph/0012099}\hbox{December 2000} }} 
 
\title{Effects from the charm scale in   
        $K^{+} \to \pi^{+} \nu \bar{\nu}$} 
\author{Adam F.~Falk and Adam Lewandowski} 
\address{Department of Physics and Astronomy, 
   The Johns Hopkins University \\ 
   3400 North Charles Street,  Baltimore, Maryland 21218 USA} 
\author{Alexey A.~Petrov} 
\address{Newman Laboratory of Nuclear Studies, Cornell 
   University\\ Ithaca, New York 14853 USA} 
\maketitle 
\thispagestyle{empty} 
\setcounter{page}{0} 
\begin{abstract}%
We consider contributions to the rare decay $K^{+} \to\pi^{+} 
\nu \bar{\nu}$ which become nonlocal at the charm scale.  
Compared to the leading term, such 
amplitudes are suppressed by powers of $m_K^2/m_c^2$ 
and could potentially give corrections at the level of 
$15\%$.  We compute the leading coefficients of the 
subleading dimension eight operators in the effective theory 
below the charm mass.  The matrix elements of these operators 
cannot all be calculated from first principles and some must 
be modeled.  We find that these contributions are likely to be 
small, but the estimate is sufficiently uncertain that the 
result may be as large as the existing theoretical 
uncertainty from other sources. 
\end{abstract} 

\newpage 
 
The search for New Physics relies on experimentally accessible 
quantities whose Standard Model values can be predicted 
accurately and reliably.  This task is often complicated by 
nonperturbative hadronic physics, especially when one is 
interested in the parameters of the 
Cabibbo-Kobayashi-Maskawa (CKM) matrix.  To make progress, 
it is important to find processes where symmetry can be used to
treat low energy QCD effects in a controlled and systematic way.  
One of these is the rare decay $K^+\to\pi^+\nu\bar\nu$. 
This process is an example of a neutral current $\Delta S = 1$
transition, which  in the Standard Model can occur only via one-loop
diagrams.

The leading contributions to the effective Hamiltonian for this 
decay are given by 
\begin{equation}\label{Hefflowest}
   {\cal H}_{eff} = \frac{G_{F}}{\sqrt{2}} \,
   \frac{\alpha}{2 \pi \sin^{2} \Theta_{W}} \sum_{l} 
   \left( V_{ts}^{*} V_{td} X(x_{t}) + V_{cs}^{*} V_{cd} 
   X_{NL}^{l}(x_{c}) \right) \bar{s}\gamma^\nu(1-\gamma^5) d\,
   \bar\nu_{l}\gamma_\nu(1-\gamma^5)\nu_{l}\,, 
\end{equation} 
where the index $l=e,\mu,\tau$ denotes the lepton flavor.  The 
coefficient $X(x_{t})$ arises from the top quark loop and is 
independent of lepton flavor.  It is dominated by calculable 
high energy physics, and has been computed to 
${\cal O}(\alpha_{s})$~\cite{toppart}.  Because it grows as 
$m_t^2$, it is large and gives the leading contribution to 
the decay rate.  If this were the sole contribution, the 
measurement of $K^+\to\pi^+\nu\bar\nu$ would yield a direct 
determination of the combination of CKM parameters 
$|V_{ts}^{*} V_{td}|$~\cite{buchalla1}.  However, due to the 
smallness of $V_{ts}^{*} V_{td}$ compared to
$V_{cs}^{*} V_{cd}$, the charm contribution contained in 
the coefficient function $X_{NL}^{l}(x_c)$ 
is significant as well.  These terms have been calculated to 
next-to-leading logarithmic order~\cite{buchallacharmQCD}. 
An important source of error in the calculation comes 
from the uncertainty in the charm quark mass, on which 
$X_{NL}^{l}(x_c)$ depends.  

An important feature of the calculation is the
fact that the hadronic matrix element
$\langle\pi^+|\bar{s}\gamma^\nu(1-\gamma^5) d|K^+\rangle$ is related
via isospin to the matrix  element
$\langle\pi^0|\bar{s}\gamma^\nu(1-\gamma^5) d|K^+\rangle$  responsible
for $K^+\to\pi^{0} e^+\nu$.  This largely  eliminates the uncertainty
due to nonperturbative QCD, up to  small isospin breaking
effects~\cite{Marciano:1996wy}. However, there remain long  distance
contributions associated with penguin diagrams  containing up quarks
which can lead to on-shell intermediate  states.  Some of these have
been estimated in chiral  perturbation theory and found to be
small~\cite{longdistance}.  The perturbative contribution from virtual
up quarks is tiny,  since it is suppressed compared to the charm
contribution by 
$m_u^2/m_c^2$. 
 
Summed over neutrino species, the branching fraction for $K^{+} 
\to\pi^+\nu\bar\nu$ is given by 
\begin{equation} 
   B(K^{+} \to \pi^{+} \nu \bar{\nu}) = \kappa_{+} \left[ \left( 
   \frac{{\rm Im}\, \xi_{t}}{\lambda^{5}}\, X(x_{t}) \right)^{2}+ 
   \left( \frac{{\rm Re}\,\xi_{c}}{3 \lambda^{5}}\, \sum_{l} 
   X_{NL}^{l}(x_c) + \frac{{\rm Re}\, \xi_{t}}{\lambda^{5}}\, X(x_{t}) 
   \right)^{2} \right],
\end{equation} 
with $\lambda=\sin\theta_C \approx 0.22$ and
\begin{equation} 
  \kappa_{+}=r_{K_{+}} \frac{3 \alpha^{2} B(K^{+} \to \pi^{0} e^{+}  
  \nu)}{2 \pi^{2} \sin^{4} \Theta_{W}}\, \lambda^{8}. 
\end{equation} 
Here $\xi_{i} = V_{is}^{*} V_{id}$, and $r_{K^{+}}$ 
absorbs isospin breaking corrections to the relationship 
between the decays $K^{+} \to \pi^{0} e^{+} \nu$ and $K^{+} \to 
\pi^+\nu\bar{\nu}$ calculated in Ref.~\cite{Marciano:1996wy}.  In terms 
of the Wolfenstein parameterization of the CKM matrix~\cite{lincoln},
the branching ratio may be written as 
\begin{equation}~\label{branchingratio} 
   B(K^{+} \to \pi^{+} \nu \bar{\nu}) = 4.11 \times 10^{-11}\cdot
   A^{4} X^{2}(x_{t}) \frac{1}{\sigma} \left[ (\sigma 
   \bar{\eta})^{2}+(\rho_{0}-\bar{\rho})^{2} \right], 
\end{equation} 
with 
$$   
  \sigma =(1-\lambda^2/2)^{-2} 
  \qquad{\rm and}\qquad\rho_{0} = 1 + \delta_{c}\,,
$$
where $\delta_{c}$ absorbs the charm contribution. A measurement 
of the branching ratio then constrains the parameters 
$\bar{\rho}$ and $\bar{\eta}$, which are equal to the 
Wolfenstein parameters $\rho$ and $\eta$ up to known corrections 
of ${\cal O}(\lambda^{2})$.  The Alternate Gradient Synchrotron (AGS)
experiment E949 at  Brookhaven and the CKM collaboration at Fermilab
propose to obtain  measurements of the branching ratio for
$K^+\to\pi^+\nu\bar\nu$ at the level of 30\% and 10\%, respectively. 
The Brookhaven experiment is the successor to AGS-E787, which saw
one event in this channel~\cite{Adler2000by}.  These experimental 
prospects then fix the goal for the accuracy of the theoretical 
prediction at less than 10\%. 
 
The leading source of theoretical uncertainty is associated 
with the charm contribution.  Calculations at next-to-leading 
order in QCD yield $\delta_c=0.40\pm0.07$, where the error is 
due primarily to the uncertainty in the charm 
mass~\cite{buchallacharmQCD}. The errors 
from uncomputed terms of order 
$\alpha^2_s(m_c)$ are expected to be small.  However, the 
computation of the charm contribution relies on an operator 
product expansion which is simultaneously a series in 
$\alpha_s$ and an expansion in higher dimension operators 
suppressed by powers of $m_c$.  The operators which are of higher 
order in the $1/m_c$ expansion reflect the fact that the penguin 
loop becomes nonlocal at the relatively low scale $m_c$.  One 
might expect the leading correction from higher order terms to 
give a contribution to $\delta_c$ of relative size 
$m_K^2/m_c^2\sim15\%$, large enough to affect in a noticeable way the
extraction of $\bar\rho$ and $\bar\eta$ from the decay rate.  It is
important  either to verify or to exclude the presence of new terms of
such a  magnitude. 
 
In this note we will study the contributions of dimension 
eight operators to the decay $K^{+}\to\pi^{+}\nu \bar{\nu}$.  
We estimate the correction to $\delta_{c}$ and  comment on the
uncertainty induced.  After discussing the relevant power counting, we
present the calculation of the operator coefficients and an  estimation
of the correction to the decay rate.  We will find a small
contribution, but one that need not be negligible.
 
\begin{figure} 
   \epsfxsize 6in 
   \centerline{\epsfbox{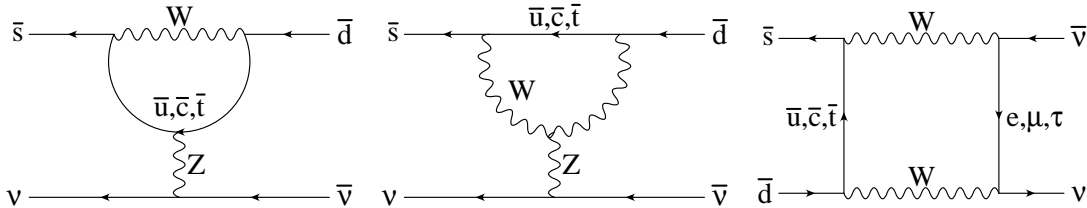}} 
   \caption{Penguin and box diagrams responsible for $K^{+} \to 
   \pi^{+} \nu \bar{\nu}$.} 
   \label{diag1} 
\end{figure}

The decay $K^{+} \to \pi^{+} \nu \bar{\nu}$ proceeds via the 
loop processes shown in Fig.~\ref{diag1}, which mediate the quark
level transition $\bar s\to\bar d\nu\bar\nu$. 
These diagrams contain both short distance and long distance 
effects, which we separate by computing the effective 
Hamiltonian density ${\cal H}_{\rm eff}$ at a low scale 
$\mu\alt1\,$GeV.  The effective Hamiltonian will receive 
corrections from the charm and top quarks, both of which have 
been integrated out of the theory, and from highly virtual up 
quarks.  Soft up quarks remain in the theory, and are 
responsible for long distance corrections. 
 
We construct the effective Hamiltonian with an operator 
product expansion.  At leading order, the operator in ${\cal 
H}_{\rm eff}$ which contributes to the decay is of dimension 
six, 
\begin{equation} 
   O^{(6)} = \bar{s} \gamma^{\nu} (1-\gamma^{5}) d \,\bar{\nu} 
   \gamma_{\nu} (1-\gamma^{5}) \nu. 
\end{equation} 
The $t$ quark contribution to the coefficient of this operator is 
obtained by evaluating the diagrams in Fig.~\ref{diag1} at the scale 
$\mu=M_W\approx m_t$ and matching on to the effective theory below
this  scale.  At the same time, the $W$ and $Z$ are integrated out of 
the theory, producing four-fermion operators involving up and 
charm quarks as well.  The charm contribution to the operator is 
then obtained by evaluating the diagrams contributing to the 
decay at $\mu= m_{c}$.  These diagrams look like those in 
Fig.~\ref{diag1}, but with the $W$ and $Z$ propagators replaced by
local interactions. 
 
Dimensional analysis indicates that the coefficient of the dimension
six operator $O^{(6)}$ scales as $1/M_{W}^{2}$.  The diagrams in
Fig.~\ref{diag1} are  quadratically divergent in the effective theory
below $M_{W}$, and scale as $\Lambda^2/M_W^4$, where $\Lambda\sim M_W$
is an ultraviolet cutoff.  The Glashow-Iliopoulos-Maiani (GIM)
mechanism ensures that this leading divergence cancels, since it is
independent of the mass $m_q$ of the virtual quark.  The consequence is
that the coefficient of $O^{(6)}$ actually scales as
$m_{q}^{2}/M_{W}^{4}$.  In terms of the Wolfenstein  parameter
$\lambda$, the top coefficient has strength 
$\lambda^{5} m_{t}^{2}/M_{W}^{4}$ and the charm coefficient has 
strength $\lambda m_{c}^{2}/M_{W}^{4}$.  The top contribution is
significant because of the large top mass, since $\lambda^4
m_t^2/m_c^2$ is of order 10. 
 
For the purpose of power counting, the operators of dimension eight
scale as
\begin{equation} 
   O^{(8)} \sim  M_{K}^{2} O^{(6)}\,, 
\end{equation} 
appearing with generic coefficient $C_{(8)}$.  Dimensionally,
$C_{(8)}$ is proportional to  $1/M_{W}^{4}$.  The top contribution to
$C_{(8)}O^{(8)}$ is suppressed by $M_K^2/m_t^2$ relative to its
contribution to $O^{(6)}$, leading to an overall strength of order
$\lambda^{5} M_{K}^{2}/M_{W}^{4}$.  The corresponding suppression for
charm is only $M_K^2/m_c^2$, so the overall contribution of charm
to $C_{(8)}O^{(8)}$ scales as $\lambda M_{K}^{2}/M_{W}^{4}$.  Note
that  there is now no relative enhancement from the large top  mass, so
the top contribution to $C_{(8)}$ is suppressed relative to that of
charm by $\lambda^4$ and can be neglected.  Furthermore, the 
contributions in question are independent of $m_q$, so they cancel by
the GIM mechanism when the up contribution is included. 
 
\begin{figure} 
   \epsfysize 1.5in 
   \centerline{\epsfbox{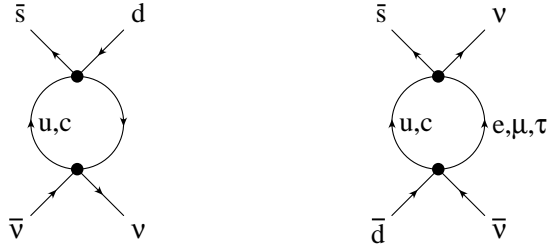}} 
   \caption{Diagrams leading to operators of dimension eight.} 
   \label{diag2} 
\end{figure}

However, the GIM cancellation is manifest in Feynman diagrams only for
contributions which are perturbatively calculable.  The long distance
contributions involving soft up quarks will differ by factors of order
one from their perturbative representations.  For these parts of the
diagrams, which scale as
$1/M_W^4$, the GIM cancellation is ineffective.  Such long distance
contributions have been considered  elsewhere~\cite{longdistance}, and
estimated to be small.  The GIM cancellation is also  spoiled by
logarithmic contributions proportional to
$(M_K^2/M_W^4)\ln{(m_c^2/M_K^2)}$.  Such terms may be generated by  the
running of ${\cal H}_{\rm eff}$ between the scale $m_c$ and  the low
energy scale $\mu\alt 1\,$GeV.  This is not a large logarithm,
numerically, but it allows us nonetheless to identify a GIM violating
contribution to ${\cal H}_{\rm eff}$.  This term, which is
generated by  intermediate up quarks as shown in Fig.~\ref{diag2},
is of  the same power-counting size as the long distance 
contribution.  But because the perturbative description of the 
long-distance part is inaccurate, there is no reason to  expect the
GIM cancellation to be restored when it is included. 
 
The purpose of this paper is to compute the corrections 
to $K^+\to\pi^+\nu\bar\nu$ of order 
$(M_K^2/M_W^4)\ln{(m_c^2/M_K^2)}$.  These contributions 
are well defined, and it is important, in light of the 
experimental situation discussed above, to determine whether 
they introduce a theoretical uncertainty at a level 
competitive with the uncertainty due to $m_c$.  Note that pure 
power counting arguments permit a relative contribution to 
$\delta_c$ of the order of $(m_K^2/m_c^2)\ln(m_c^2/\mu^2)$, which 
could be as large as $20\%$, depending on the value chosen for the 
hadronic scale $\mu$. 
 
We will study the effective Hamiltonian of dimension 
eight operators, at leading order in $\alpha_s$.  This 
Hamiltonian receives logarithmically enhanced contributions 
from the up quark loops in Fig.~\ref{diag2}.  We also must consider the
matching corrections at the scale $m_c\approx m_\tau$, when the tau
lepton is integrated out of the theory.  Because the matching function
$F(m_c/m_\tau)$ cannot  be approximated by an expansion in
$m_c/m_\tau$, the  combination $[F(m_c/m_\tau)-F(m_u/m_\tau)]$ is a GIM
violating finite matching correction which also must be included. 
 
The effective Hamiltonian density at the scale $\mu$ takes the form 
\begin{equation} 
   {\cal H}_{\rm eff} = \sum_{l, i} C_{i}^{l}(\mu) 
   O_{i}^{l}(\mu)\,,
\end{equation} 
where $l$ denotes lepton flavor. As it turns out, there will be 
two dimension eight operators generated in the theory below 
$m_c$,
\begin{eqnarray} 
   O_{1}^{l} &=& \bar{s} \gamma^{\nu} (1-\gamma^{5}) d \, (i 
   \partial)^{2} \left[ \bar{\nu_{l}} \gamma_{\nu} (1-\gamma^{5}) 
   \nu_{l} \right], \nonumber \\ O_{2}^{l} &=& 
   \bar{s} \gamma^{\nu} (1-\gamma^{5})(i 
   {D})^{2} d \, \bar{\nu_{l}} \gamma_{\nu} 
   (1-\gamma^{5}) \nu_{l} + 
   2 \bar{s} \gamma^{\nu} (1-\gamma^{5}) (i 
   {D^{\mu}}) d \, \bar{\nu_{l}}\gamma_{\nu} 
   (1-\gamma^{5}) (i 
   {\partial_{\mu}}) \nu_{l}\nonumber  \\ 
   & & \quad \mbox{} + \bar{s} \gamma^{\nu} (1-\gamma^{5}) d \, 
   \bar{\nu_{l}} 
   \gamma_{\nu} (1-\gamma^{5}) 
   (i {\partial})^{2} \nu_{l}\,.
\end{eqnarray} 
The first of these operators does not receive any logarithmic 
QCD corrections below the scale $m_c$, because it is proportional to
a current which is partially conserved.  The second does,
but  we will not include higher order corrections of relative order 
$\alpha_s\ln(m_c/\mu)$.  Note that this is not inconsistent with
resumming terms of order $\alpha_s^n\ln^n(M_W/m_c)$. 
 
\begin{figure} 
   \epsfysize 1.5in 
   \centerline{\epsfbox{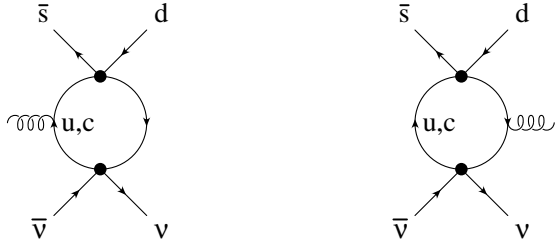}} 
   \caption{Diagrams which could lead to an operator with a 
   gluon field strength.} 
   \label{diag3} 
\end{figure}

The operators $O_{1,2}^l$ are generated by the diagrams in 
Fig.~\ref{diag2}.  In principle, one might have expected the diagrams
in Fig.~\ref{diag3} to generate additional operators with a gluon field
strength, such as 
\begin{equation} 
   O_{3}^{l} = \bar{s} \gamma^{\nu} \sigma^{\alpha \rho} 
   G_{\alpha \rho} (1-\gamma^{5}) d \, \bar{\nu_{l}}  \gamma_{\nu} 
   (1-\gamma^{5}) \nu_{l}\,.
\end{equation}
However, it turns out that contributions to all such operators 
cancel. 
 
The operators of dimension six that will induce $O_{1,2}^l$ in 
${\cal H}_{\rm eff}$ are 
\begin{eqnarray} 
   O_{4} &=& \bar{s} \gamma^{\nu} (1-\gamma^{5}) d\, \bar{u} 
   \gamma_{\nu} (1-\gamma^{5}) u\,, \nonumber \\ O_{5} &=& \bar{s} 
   \gamma^{\nu} (1-\gamma^{5}) u\,\bar{u} \gamma_{\nu} 
   (1-\gamma^{5}) d\,, \nonumber \\ O_{6}^{l} &=& \bar u\gamma_{\nu}
   \left( - \textstyle{\frac{4}{3}} 
   \sin^{2} \theta_{W} (1+\gamma^{5}) + ( 1- 
   \textstyle{\frac{4}{3}} 
   \sin^{2} \theta_{W}) (1-\gamma^{5}) \right) u
   \,\bar{\nu_{l}}\gamma^{\nu} 
   (1-\gamma^{5}) \nu_{l}\,,\nonumber  \\ 
   O_{7}^{l} &=& \bar{s} \gamma^{\nu} (1-\gamma^{5}) u\, 
   \bar{\nu_{l}} \gamma_{\nu} (1-\gamma^{5}) l\,, \nonumber \\
   O_{8}^{l} &=& 
   \bar{u} \gamma^{\nu} (1-\gamma^{5}) d\, \bar{l} \gamma_{\nu} 
   (1-\gamma^{5} ) \nu_{l}\,. 
\end{eqnarray} 
The operator $O_6^l$ comes from virtual $Z$ exchange, the others from
$W$ exchange.  The renormalization group equations for
$O_{1}^{l}$  and $O_{2}^{l}$ are 
\begin{eqnarray} \label{C2renorm}
   \mu \frac{d C_{1}^{l}}{d \mu} &=&= \gamma_{1} C_{4} C_{6}^{l}+ 
   \gamma_{1}' C_{5} C_{6}^{l}\,, \nonumber\\
   \mu \frac{d C_{2}^{l}}{d \mu} &=&= \gamma_{2} C_{7}^{l} C_{8}^{l} 
   + \sum_{j} \gamma_{2j} C_{j}^{l}\,. 
\end{eqnarray} 
The anomalous dimensions $\gamma_1$, $\gamma'_1$ and $\gamma_2$ are
of order one.  The matrix $\gamma_{2j}$ is of order $\alpha_s$ and
comes from QCD running below $m_c$; it will not be included in our
analysis.
 
Computing the diagrams in Fig.~\ref{diag2} and solving the 
renormalization group equations, we find the 
coefficients at the scale $\mu$, 
\begin{eqnarray} 
   C_{1}^{e,\mu,\tau} &=& \frac{c_{0} }{6 M_{W}^{2}} 
   (1-\textstyle{\frac{4}{3}} \sin^{2} \Theta_{W}) G(\alpha_{s}) 
   \log(\mu/m_c)\,, \nonumber
   \\ 
   C_{2}^{e,\mu} & = & - \frac{c_{0}}{M_{W}^{2}} \log(\mu/m_c)\,, 
    \\ 
   C_{2}^{\tau}  &=& - \frac{c_{0}}{4 M_{W}^{2}} f ( 
   m_c^2/m_\tau^2)\,, \nonumber
\end{eqnarray} 
where 
\begin{eqnarray} 
   c_{0} &=& \frac{G_{F}}{\sqrt{2}}\, \frac{\alpha}{2 \pi 
   \sin^{2}\theta_{W}} \,V_{cs}^{*} V_{cd}\,, 
    \nonumber\\ 
   f(x) &=& \left( \frac{6x-2}{(x-1)^{3}}-2 \right) \log x - 
   \frac{4 x}{(x-1)^2}\,, 
    \\ 
   G(\alpha_{s}) &=& 2 \left( 
   \frac{\alpha_{s}(m_{c})}{\alpha_{s}(m_{b})} \right)^{-6/25} 
   \left(\frac{\alpha_{s}(m_{b})}{\alpha_{s}(M_{W})} 
   \right)^{-6/23} - \left( 
   \frac{\alpha_{s}(m_{c})}{\alpha_{s}(m_{b})}  \right)^{12/25} 
   \left( \frac{\alpha_{s}(m_{b})}{\alpha_{s}(M_{W})} 
   \right)^{12/23}, \nonumber
\end{eqnarray} 
and we have used $V^{*}_{us} V_{ud} \approx -V_{cs}^{*} V_{cd}$. Taking the 
values $m_{c}=1.3\,$GeV, $m_{b}=4.5\,$GeV, $\Lambda_{\overline{\rm
MS}}=0.35\,$GeV and $\mu = m_c/2$, we find
\begin{equation}
  C_{1}^{e,\mu,\tau}=0.05\cdot c_0/M_W^2\,,\qquad
  C_{2}^{e,\mu}=0.69\cdot c_0/M_W^2\,,\qquad 
  C_{2}^{\tau}=0.28\cdot c_0/M_W^2\,.
\end{equation}
By comparison, the coefficient of the leading charm contribution in
Eq.~(\ref{Hefflowest}) is given by $X_{NL}^l(x_c)\,c_0$, which is
$4.0\,m_c^2\cdot c_0/M_W^2$ for $l=e,\mu$ and $2.7\,m_c^2\cdot
c_0/M_W^2$ for $l=\tau$.

To compute the contribution to the decay rate, we also need  
the matrix elements $\langle\pi^+\nu_l\bar\nu_l|\,O_{1,2}^l\,| 
K^+\rangle$.  The leading relative corrections come from the
interference of $O_{1,2}$ with $O^{(6)}$ and depend on the ratios
\begin{equation}
   R_{i} = \frac{{\rm Re}\,\int d[{\rm P.S.}]\,|
  \langle\pi^+\nu_l\bar\nu_l|\,O_i^l\,| K^+\rangle^*
  \langle \pi^{+} \nu \bar{\nu} | O^{(6)} |
   K^{+} \rangle}{\int 
   d[{\rm P.S.}]\, |\langle \pi^{+} \nu \bar{\nu} | O^{(6)} |
   K^{+} \rangle|^2}\,. 
\end{equation}
The matrix element of the operator $O_{1}$ is easy to
calculate, since it depends only on the lepton momenta.  The leptons
are treated perturbatively, so the hadronic dependence of the matrix
element of $O_{1}$ is the same as that of $O^{(6)}$.  We then find
\begin{equation}
   R_{1} =\langle(p_{\nu} + p_{\bar{\nu}})^{2}\rangle=(180\,{\rm
   MeV})^2\,. 
\end{equation}

Unfortunately, the matrix element of $O_{2}$ cannot be calculated
analytically, since it involves the gluon field through the covariant
derivative acting on the down quark.  We are forced to rely instead
on model dependent estimates, which are notoriously unreliable.  One
ansatz would be to take
\begin{equation} \label{naiveestimate}
   \langle \pi^{+} \nu \bar{\nu} | O_{2} | K^{+} \rangle_{\mu} 
   \approx\mu^{2}\, \langle \pi^{+} \nu \bar{\nu} | O^{(6)} | K^{+} 
   \rangle_{\mu}, 
\end{equation} 
or $R_2\approx\mu^2\sim(650\,{\rm MeV})^2$.  Another would be to
neglect the gluon field and model the matrix element as 
\begin{equation}~\label{matrixelement} 
   \langle \pi^{+} \nu \bar{\nu} | O_{2} | K^{+} \rangle = 
   (p_{\pi} + p_{\bar{\nu}})^{2}\, \langle \pi^{+} \nu \bar{\nu} | 
   Q^{(6)} | K^{+} \rangle, 
\end{equation}
in which case
\begin{equation}
   R_{2}\approx\langle(p_{\pi} + p_{\bar{\nu}})^{2}\rangle=(340\,{\rm
   MeV})^2\,. 
\end{equation}
Of course, neither of these guesses need be correct within better
than an order of magnitude.  Fortunately, lattice QCD methods are 
advancing quickly, to the point that a true unquenched lattice 
calculation of this matrix element may soon be feasible.  For now, we
will take these two crude guesses to bracket roughly the actual value of
$R_2$.
 
We now write the branching fraction for $K^+\to\pi^+\nu\bar\nu$ as in
Eq.~(\ref{branchingratio}), with 
\begin{equation} 
   \rho_{0} = 1 + \delta_{c}(1+\delta_{8})\,, 
\end{equation} 
where $\delta_8$ is the new term which we are computing.  Summed over
lepton species, the contribution of charm at  dimension six is given by 
\begin{equation} 
   \delta_c={P_0(x_c)\over A^2 X(x_t)}=  
   \frac{1}{3\lambda^{4}}\sum_{l}X_{NL}^{l}(x_c)
   \cdot {1\over A^2 X(x_t)}\,. 
\end{equation} 
A next to leading order analysis yield $P_{0} = 0.42 \pm 
0.06$, where the error arises in large part from the uncertainty 
in the charm quark mass~\cite{buchallacharmQCD}.  This value of
$P_{0}$ gives $\delta_{c} = 0.40 \pm 0.07$, where we
use $X(x_{t})=1.53 \pm 0.01$ and $A = 0.83 \pm 0.06$. 
The fractional correction due to dimension eight operators is then
\begin{equation} 
   \delta_{8} =\frac{1}{3 P_{0}\lambda^{4}}\, \sum_{l}\left( 
   C_{1}^{l} 
   R_{1}+C_{2}^{l} R_{2} \right)=\delta_8^{(1)}+\delta_8^{(2)}\,. 
\end{equation} 
The first term, for which the matrix element is calculable, is
negligible in size: with our choice of inputs,
$\delta_8^{(1)}=5.6\times10^{-5}$.  The second, highly uncertain,
term is much bigger, with $\delta_8^{(2)}$ between 1\% and 5\% for our
adopted range for $R_2$.  On the one hand, even $\delta_8$ as large as
5\% is somewhat below the existing uncertainty on $\delta_c$ from the
value of $m_c$.  On the other, if our ``upper limit'' on the matrix
element of $O_2^l$ were too small by even a factor of two, which need
not be unlikely, these contributions would have a significant effect on
the extraction of CKM parameters from the branching fraction.

We have made a number of approximations in obtaining these results. 
A potentially important one, within the perturbative calculation, is
that we have neglected QCD  running below $m_c$.  We could include
these QCD corrections for $O_{1}$ simply by incorporating the known
running of the  coefficients $C_{4}$ and $C_{5}$; doing so decreases
$C_{1}$ by  a factor of two.  However, the running of $O_{2}$ is not
equally trivial, since $O_2$ itself is renormalized in QCD. In view of
the large uncertainty in the matrix element of  $O_2$, including these
QCD corrections would not at this time increase the reliability of our
prediction. 
 
Of course, the key uncertainty arises not from QCD perturbation theory 
but from the actual value of $ \langle \pi^{+} \nu \bar{\nu} |\, 
O_{2}\, | K^{+} \rangle$.  Only a realistic lattice computation will
settle the matter.  We would argue, in fact, that such a calculation
is really required for one to be confident that the effects we have
considered do not  spoil the extraction of CKM matrix elements from the
proposed  experiments on $K^+\to\pi^+\nu\bar\nu$.  This is not the
only case where higher dimensional operators can play an interesting
role in kaon decays~\cite{buchalla,Cirigliano:2000ev}.
 
In summary, we have computed the dominant contribution to the 
coefficients of dimension eight operators contributing to the 
decay $K^{+} \to \pi^{+} \nu \bar{\nu}$.  Our best estimate is 
that this represents a correction of no more than 5\% to the leading
charm contribution to the decay.  However, our ignorance of relevant
hadronic matrix elements leaves open the possibility that these
contributions could represent an uncertainty as large as or larger than
that due to the charm quark mass.  A lattice calculation of
nonperturbative  corrections, and to a lesser extent the
inclusion of perturbative QCD corrections below the charm scale, will
be indispensable to reducing this uncertainty  before the planned
experiments begin to take their data. 
 
\acknowledgments 
 
We are grateful to Mark Wise for discussions.  A.F.~and A.L.~are 
supported in part by the United States National Science 
Foundation under Grant No.~PHY-9404057 and by the United States 
Department of Energy under Outstanding Junior Investigator 
Award No.~DE-FG02-94ER40869.  A.P. is supported in part by the 
National Science Foundation.  A.F.~is a Cottrell Scholar of the 
Research Corporation.


\begin{references} 
 
\bibitem{toppart} T. Inami and C.S. Lim, Prog. Theor. Phys 
{\bf 65} 297 (1981);\\ G. Buchalla and A. Buras, Nucl. Phys. 
{\bf B398}, 285 (1993);\\ G. Buchalla and A. Buras, Nucl. Phys. 
{\bf B400}, 255 (1993);\\ M. Misiak and J. Urban, Phys. Lett. 
{\bf B451}, 161 (1999). 
 
\bibitem{buchalla1} G. Buchalla and A.J. Buras, Nucl. Phys. 
{\bf B548}, 309-327 (1999). 
 
\bibitem{buchallacharmQCD} G. Buchalla and A. Buras, Nucl. Phys. 
{\bf B412}, 106 (1994). 
 
\bibitem{Marciano:1996wy}
W.J. Marciano and Z. Parsa,
Phys.\ Rev.\  {\bf D53}, 1 (1996).

\bibitem{lincoln}
L. Wolfenstein, Phys.\ Rev.\ Lett. {\bf 51}, 1945 (1983).

\bibitem{Adler:2000by}
S. Adler {\it et al.} [E787 Collaboration],
Phys.\ Rev.\ Lett.\  {\bf 84}, 3768 (2000).

\bibitem{longdistance} M. Lu and M.B. Wise, 
Phys. Lett. {\bf B324},  461 (1994);\\ C.Q. Geng, I.J. Hsu and 
Y.C. Lin, Phys. Lett. {\bf B355}, 569 (1995);\\ C.Q. Geng, I.J. 
Hsu and C.W. Wang, Prog. Theor. Phys. {\bf 101}, 937 (1999). 
 
\bibitem{buchalla}
G. Buchalla and G. Isidori, Phys. Lett. {\bf
B440}, 170 (1998).

\bibitem{Cirigliano:2000ev} 
V. Cirigliano, J.F. Donoghue and E. Golowich,
JHEP {\bf 0010}, 048 (2000).

\end{references}
\end{document}